\documentstyle{l-aa}                   
\newif\ifnofig\nofigfalse   
\newif\ifthisfig
\ifnofig\else\input{psfig}\psfigurepath{figure/}\fi

\def\setfigdefault{\ifnofig\thisfigfalse\else\thisfigtrue\fi}
\setfigdefault
\newcommand{\bp}{$\beta\:$Pictoris}

\begin{document}
\thesaurus{08 (08.09.2 $\beta\:$Pic; 08.03.4; 08.16.2; 03.13.4)}
\title{Dust distribution in
disks supplied by small bodies:\\
Is the \bp\ disk a gigantic multi-cometary tail?}

\author{
A. Lecavelier des Etangs \inst{1} \and
A. Vidal-Madjar \inst{1} \and
R. Ferlet \inst{1} }

\offprints{A. Lecavelier des Etangs}
\institute{
Institut d'Astrophysique de Paris, 98 Boulevard Arago,
F-75014 Paris, France
}

\date{Received; accepted}

\maketitle

\markboth{A. Lecavelier des Etangs et al.: Dust distribution in
disks supplied by small bodies.}{}

\begin{abstract}
We have evaluated the spatial distribution of dust in disks supplied by
colliding or evaporating bodies spread in a small belt. The
gradient of the distribution
of dust expelled by radiation pressure is generally steeper than
the one observed around \bp. It follows that, to generate the \bp\ disk
distribution in such a way, one should extract relatively more small particles
in the production process: thanks to radiation forces, these smaller particles
have large eccentricities and thus could be seen at very large distances from
their injection place. The evaporation process of comet-like bodies moving
slowly inwards may yield the necessary particle size distribution.

It is interesting to note that this dynamical description is able to account
for the main observational properties of the \bp\ disk, such as the
power law distribution
with possibly a change in slope, the asymmetry at large distances, and
the total
mass, all of which have remained unexplained up to now.
If confirmed, this scenario may
indicate that the \bp\ disk could be looked on
as a gigantic multi-cometary tail
with its natural constituents: gas and dust.


\end{abstract}

\section{Introduction}
\label{intro}

Since the discovery of the dusty disk around \bp\ (Aumann 1984, Smith \&
Terrile 1984), it has been the subject of extensive investigations with all
possible observational techniques: visible and IR imagery (Smith \& Terrile
1987, Paresce \& Burrows 1987, Lecavelier des Etangs et al. 1993, Golimowski et
al. 1993, Lagage \& Pantin 1994), spectroscopy (Vidal-Madjar et al. 1994,
Vidal-Madjar \& Ferlet 1995 and references therein) and recently photometric
variations (Lecavelier des Etangs et al. 1995). In particular, the stellar
light scattered by dust particles has been well observed by several authors for
many years to follow a power law distribution with a slope in the range -3.6
(Artymowicz et al. 1989, Lecavelier des Etangs et al. 1993) -4.3 (Smith \&
Terrile 1984). However, a convincing explanation for this scattered light
distribution is still needed.

The dust distribution in the \bp\ disk may be a crucial issue for understanding
its origin. We present here a precise dynamical model which will be compared
with the observations for the first time in a quantitative manner.

Another key point revealed by the spectroscopic studies of the gaseous
counterpart of the \bp\ disk is the existence of orbiting kilometer size bodies
which sometimes fall onto the star, giving rise to the observed
spectroscopic redshifted signatures due to ejected gas,
and undoubtedly to dust particles. Already
in 1984, Weissman (1984) questioned whether the material around Vega-like stars
is of cometary or asteroidal origin and suggested that particles could be
continually supplied by sublimation or from collisions between larger bodies.

The time scale evaluation over which dust particles are eliminated by the
Poynting-Robertson effect or collisions (Backman \& Paresce 1993) leads to
durations much shorter than the estimated age of \bp\ ($\sim 2\cdot 10^{8}$
years according to Paresce, 1991). This simple comparison suggests convincingly
that the \bp\ disk is not a remnant of planetary formation but on the contrary
must be continually replenished by secondary sources like evaporation or
collision of small bodies. This seems in fact a very natural hypothesis (see
for example Zuckerman \& Becklin 1993): the presence of small bodies is a
consequence of planetary systems formation (Lissauer 1993).

Furthermore, CO absorptions have been recently observed towards \bp\
(Vidal-Madjar et al. 1994) with the Hubble Space Telescope. The very presence
of CO may also require a permanent source provided by evaporation of comet-like
bodies.

Here we present a new argument: This replenishment is able to reproduce
the main characteristics of the \bp\ dust disk, and in particular
the spatial distribution of dust at large distances. In detail, it gives
 possible explanations for the following unexplained issues:

\begin{enumerate}
\item The gradient of the scattered light follows a relatively well-known
but unexplained power law (e.g. Golimowski et al. 1993).
\item The distribution at large distances is obviously not axisymmetric
(Smith \& Terrile 1987, Kallas \& Jewitt 1995).
\item The central part of the disk is relatively clear of dust
(Backman et al. 1992, Lagage \& Pantin 1994).
\item The disk seems to be a ``wedge'' disk: the thickness increases
with radius (Backman \& Paresce 1993, Lecavelier des Etangs et al. 1993).
\item The slope of the scattered light distribution
changes abruptly at about 100~AU from the star.
If confirmed, this fact remains unexplained
(Artymowicz et al. 1990, Golimowski et al. 1993).
\end{enumerate}

Furthermore, the possible explanation of these observational facts could
give new understanding to the following questions:
\begin{itemize}
\item a) What is the mass of the dust disk, since when extrapolated towards
infinite distances the disk mass diverges (Artymowicz et al. 1989)?
\item b) Are there connections between the dust and the gas disks ?
\item c) Can the asymmetry in the observed dust disk be connected with the
asymmetry in the longitude of periastron of the comets in the
Falling-Evaporating-Bodies (FEB) model proposed to explain the
redshifted spectroscopic events
(e.g. Beust et al. 1991)?
\end{itemize}

Therefore, we will consider here a model in which the disk is replenished by a
group of kilometer-size bodies (Section~\ref{model}). Numerical calculations of
the spatial distribution of dust are given in Section~\ref{numerical}. We shall
discuss in Section~\ref{bp disk} the \bp\ disk in such a scenario and summarize
the results in Section~\ref{conclusion}.

\section{The model}
\label{model}

\subsection{Basic concept}

We assume that the dust is produced by cometary or asteroid-like
bodies which create particles through mutual collisions or evaporation
processes. The main assumption is thus that these particles have small initial
velocities relative to the parent bodies: less than 1 km/s like in cometary
production (Gombosi et al. 1985, Sekanina 1987) or in "Chiron burst"
(Luu \& Jewitt 1990). Thus, these velocities
can be considered as negligible in comparison to the orbital velocities.
In our model, this relative velocity is fixed at zero.
However, as soon as a dust particle is ejected, it is perturbed by the
radiation pressure; its orbit is different from the parent body one
and tangent
at the point where it was injected
(Burns et al. 1979). For example, if a
parent body in a circular orbit with a semi-major axis $a_0$ produces
a particle with a ratio $\beta $
of the radiation force to the gravitational force,
the particle orbit has
a semi-major axis $a_{\beta}=a_0(1-\beta )/(1-2\beta )$ and
an eccentricity $e_{\beta}=\beta /(1-\beta )$. The periastron
is $a_{\beta}(1-e_{\beta})=a_0$ and the apoastron
$a_{\beta}(1+e_{\beta})=a_0/(1-2\beta )$.

Therefore, particles can be observed at distances from the central star much
larger than the apoastron of the parent body. Thus, local perturbations on
the distribution of the parent bodies like asymmetries, could produce
observable
signatures on the dust distribution at very large distances.

\subsection{Particle size distribution}
\label{size}

Furthermore, because the parent bodies should not produce single sized
particles,
we introduce in the calculation a size distribution. For
particles with radius $s\ge 1\mu$,
$\beta$ is correlated with the size of the particle by $\beta\sim s^{-1}$
(Artymowicz 1988). We can assume a power
law size distribution as in Solar System cometary
dust (Lien 1990 and references therein),
in collisionally replenished dust (Greenberg \& Nolan 1989),
or in interplanetary medium dust (Le Sergeant \& Lamy 1980).
Thus, if we have a size distribution $dn\propto s^{q}ds$, then we have
$dn\propto \beta^{K}d\beta$ with $K=-q-2$.

\subsection{Analytical consideration}
\label{analytical}
\subsubsection{One point production}

First, let us consider a parent body in a circular orbit
which generates a set of particles
at a given point. Then the particles with $\beta >0.5$ have hyperbolic
orbits and are ejected from the system towards the interstellar medium. The
other particles with $\beta <0.5$ follow orbits within a parabola which
is the orbit of the particles with $\beta=0.5$.

One can evaluate the surface density of particles in the asymptotic
direction of the parabola: for a given particle the true anomaly $\theta$
follows the distribution law

\begin{center}
\begin{equation}
P_{\theta}(\theta )d\theta =\frac{(1-e)^{3/2}}
                 {2\pi (1+e\cos \theta)^2}    d\theta
\end{equation}
\end{center}

In the asymptotic direction $\theta=\pi$, $P_{\theta}(\pi)\propto (1-e)^{-1/2}
\propto r^{1/2} (1+a_0/r)^{1/2}$. The probability that the particle apoastron
is between $r$ and $r+dr$ in the same direction
is $P_r(r)= \beta^K (d\beta /dr) dr$. Since
$\beta=(1-a_0/r)/2$, $P_r(r)\propto (1-a_0/r)^K r^{-2} dr$.
Thus, taking into account the scattering cross section
$\sigma\propto \beta^{-2}$ we obtain the surface
density normal to the plane of the disk
$\sigma \cdot n_s$:

\begin{center}
\begin{equation}
\sigma \cdot 2\pi r\cdot n_s(r)\propto r^{-3/2}(1-a_0/r)^K (1+a_0/r)^{1/2}
\end{equation}
\end{center}

Thus $\sigma \cdot n_s(r)_{r \rightarrow \infty}\propto r^{-2.5}$.

\subsubsection{Parent bodies in circular orbits.}
\label{circular orbits}

\begin{figure}[tbp]
\ifthisfig
\psfig{file=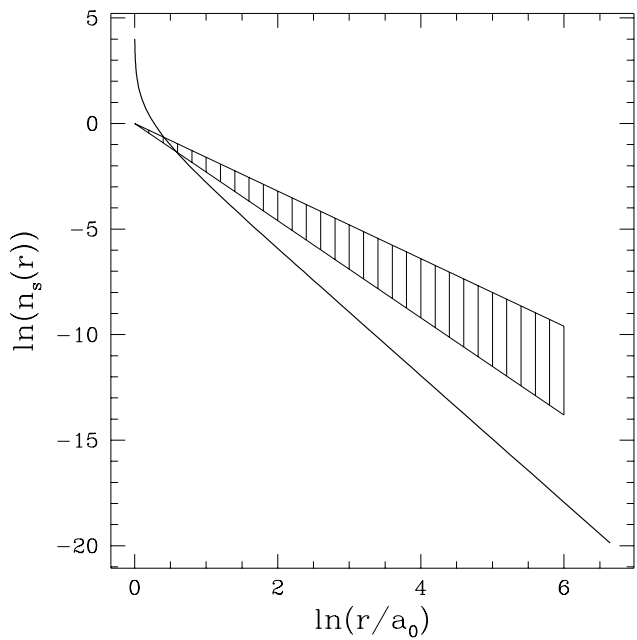,height=\columnwidth,rheight=8.5truecm}
\else\picplace{6truecm}\setfigdefault\fi
\caption[]{Plot of the surface density distribution calculated from
Eq.~\ref{circular}. The abrupt density decrease at $r\sim a_0$ is
simply due to the assumed single distance between the central star and
all the parent bodies. It is thus an artificial consequence of an
oversimplified hypothesis.\\
For comparison, the hatched zone represents the distribution range of slopes
deduced by different observers around \bp.}
\label{fig1}
\end{figure}

However, dust production can take place at every point in the
orbit of the parent body.
Here, we restrict ourselves to a parent body in circular orbit,
thus, producing an axisymmetrical disk.
Each particle has a probability $f_{\beta}(r)$ to be at a distance $r$:
$f_{\beta}(r)\propto ((1-a_0/r)(2\beta -1 +a_0/r))^{-1/2}$.
The surface
density normal to the plane of the disk is then:

\begin{center}
\begin{equation}
\sigma \cdot n_s(r)\propto \int_{(r-a_0)/2r}^{1/2}
\frac{\beta^{K-2} f_{\beta}(r) d\beta }
     {2 \pi r\int_{a_0}^{a_0/(1-2\beta)} f_{\beta}(r_1)dr_1}
\label{circular}
\end{equation}
\end{center}
This surface density follows an $r^{-3}$ law and is plotted in
Fig.~\ref{fig1}.

Moreover, the inclination of the particles is the same as the inclination of
the parent bodies. Therefore, the vertical distribution of the particles
depends on the distribution of the inclination of the parent bodies. In the
present
model the thickness of the disk is increasing with radius and the volume
density ($n_v$) is proportional to the surface density ($n_s$) divided by the
distance, i.e. $n_v(r)\propto r^{-4}$.

It has to be noted that
this conclusion is valid at distances larger than the distance of the farthest
parent bodies. Within this zone the opening angle must look smaller.

{}From the $n_v(r)\propto r^{-4}$ law and Nakano's (1990) conclusion
concerning the
connection between the volume density and the scattering light distribution
($F(r)\propto n_v(r)/r $),
one can conclude that a belt of parent bodies in circular orbits should produce
a "wedge" disk with a scattered light distribution $F(r)\propto r^{-5}$.

\section{Numerical results}
\label{numerical}
\subsection{The Monte-Carlo simulation}

The analytical model of the previous section can only
be solved for a limited number of
parameters and parent body distributions.
We have chosen to evaluate the dust distribution by a
Monte-Carlo method.
We shall consider a given family of parent bodies, a dust production
law and a particle size distribution, and we shall integrate the distribution
in the following way:

For each of the $N$ random particles, we first
randomly select a parent body in the family and its position in its
orbit; then we randomly apply the particle size distribution and
we calculate the position $r_i$ of the particle $i$ after a time $t$
where $t$ is randomly chosen between
$0$ and $t_{max}$.
Finally, we evaluate $F(r)$
the surface brightness along the midplane
of the edge-on disk projected on the sky at the projected distance $r$.
If $F_i(r)$ is the contribution of the particle $i$,
from Eq.~2 of Nakano (1990), we have
\begin{center}
\begin{equation}
F_i(r)\propto r^{-1}\int _0 ^{\pi}n_i(\Lambda)\sigma_i(\theta)d\theta
\end{equation}
\end{center}
where $\Lambda$ is the distance to the central
star and $\sigma_i(\theta)$ is the scattering phase function.
Nakano (1990) has already shown that, except if the dust
is much more steeply forward-scattering than in the Solar System,
the isotropic scattering assumption does not
change the observed slope of the scattered light
distribution, thus $\sigma_i(\theta)=\beta _i^{-2}$ with
$\sin \theta = r/\Lambda$.
The volume density is given by
$n_i(\Lambda) = \delta(\Lambda-r_i)\cdot p_{i} /r_i$ where
$p_i$ is the production rate of dust on the parent bodies,
then
$F_i(r)\propto r^{-1}\int \delta(\Lambda-r_i) p_i {\rm H}(\Lambda-r)r
                                                            d\Lambda /
               (r_i \beta _i^{2} \Lambda \sqrt{\Lambda^2-r^2})$
where
H is the Heavyside function.

Finally, the surface brightness
at a given projected distance $r$ from the star is estimated by
\begin{center}
\begin{equation}
F(r)=\sum_{i=1}^{N}F_i(r)
\propto \sum_{i=1}^{N}\frac{ {\rm H}(r_i-r)\cdot p_{i} }
                    {r_{i}^{2} \sqrt{r_i^2-r^2}\cdot \beta _{i}^{2}}
\end{equation}
\end{center}

\subsection{Results}

Results of runs with different initial conditions on the family of parent
bodies
are summarized in Tables~\ref{resultat_1} and \ref{resultat_2}.
The production rates per annulus of fixed width are given for each run. Several
configurations have been tried. The first one corresponds to the analytical
model of Sect.~\ref{circular orbits}, we find again $\alpha=5.0$, which
validates the Monte-Carlo model. In runs
\#2 and \#3, we have tried different distances and evolutionary times
which show
that the Poynting-Robertson effect does not change the results: the accuracy
for $\alpha$ is about 0.1. From \#4 to \#11, the parent bodies are in
eccentric orbits with eccentricities uniformly distributed between 0 and 0.5 or
between 0 and 1; it seems that bodies with eccentricities larger than 0.5 do
not contribute much to the distribution since a large fraction of ejected
material is then on hyperbolic orbits. In runs \#4 to \#7,
we have taken parent bodies
with a peculiar longitude of periastron ($\omega$),
namely 155$^o$. This value has been selected according
to the Beust et al. (1991) modelisation of the Falling-Evaporating-Bodies,
the so-called FEB scenario needed to reproduce the spectroscopic redshifted
signatures. In the four last runs of Table~\ref{resultat_1}, $\omega$ was
uniformly distributed between 0$^o$ and 360$^o$.

{}From all the results in
Table~\ref{resultat_1}, it can be seen that the slope $\alpha$ is always
greater than or equal to 5.0. In Table~\ref{resultat_2}, we test other parent
body
distributions in which the dust observed at large distances still is
produced close to the star. Incidentally, another solution would be to assume
that the disk is produced by a distribution of asteroids up to 1000~AU from the
star. However this solution does not allow us to explain the observed
asymmetries.

Runs \# 12 to \# 15 take into account the Epstein gas drag with a gas
density $\rho_{gas}=\rho _0 (100\ {\rm AU}/r)$  and temperature $T_{gas}=20$~K.
Runs \# 16 and \# 17 are with dust size distribution characterized by
$K=5-10$.

\begin{table*}
\caption{Slope $\alpha$ of the gradient of brightness in disks with different
conditions for the parents bodies.
In all these runs we assumed N=10000 and K=1.5. Poynting-Robertson drag
is taken into account but there is no gas drag.
For runs \#~4 to \#~7, the two slopes are for the both sides of the
asymmetrical disk seen from the Earth.}
\label{resultat_1}
\begin{tabular}{|c|cccc|c|c|}
\hline
Run&&Parent Bodies Parameters &&&$t_{max}$ &Result: $\alpha$\\

$\#$ & semi-major axis (AU) & Eccentricity & $\omega$ & Production rate &
(years)&$F(r)\propto r^{-\alpha}$     \\
\hline
1 & 20 & 0. & - & $p=const$ & 10000 				& 5.0 \\
2 & 2 & 0. & - & $p=const$ & 10000 				& 4.9 \\
3 & 20 & 0. & - & $p=const$ & 100000 				& 5.0 \\
4 & 20-30 & 0.-0.5 & 155$^o$ & $p=const$ & 10000 		& 5.2-5.6 \\
5 & 20-30 & 0.-1. & 155$^o$ & $p=const$ & 10000 		& 5.2-5.7 \\
6 & 20-30 & 0.-0.5 & 155$^o$ & $p\propto r^{-3}$ & 10000 	& 5.3-5.5 \\
7 & 20-30 & 0.-1. & 155$^o$ & $p\propto r^{-3}$ & 10000 	& 5.3-5.6 \\
8 & 20-30 & 0.-0.5 & 0-360$^o$ & $p=const$ & 10000 		& 5.2 \\
9 & 20-30 & 0.-1. & 0-360$^o$ & $p=const$ & 10000 		& 5.3 \\
10 & 20-30 & 0.-0.5 & 0-360$^o$ & $p\propto r^{-3}$ & 10000 	& 5.2 \\
11 & 20-30 & 0.-1. & 0-360$^o$ & $p\propto r^{-3}$ & 10000 	& 5.4 \\
\hline
\end{tabular}
\end{table*}

\begin{table*}
\caption{Same as the previous table, with different $K$; gas drag is
taken into account. The parent bodies have semi-major axis $a=$20~AU and
eccentricity $e=0$. $t_{max}$=10000. For the runs \# 16 the slope
gradually changes from $\alpha=3.7$ at 20 AU to $\alpha=4.7$ at 800 AU; and for
run \# 17 from 2.9 to 4.4 respectively.}
\label{resultat_2}
\begin{center}
\begin{tabular}{|c|c|c|c|}
\hline
Run  & $\beta$ distribution:K& gas density at 100 AU & Result: $\alpha$ \\
$\#$ & $dn=\beta ^{K}d\beta$ & (cm$^{-3}$)           & $F(r)\propto
r^{-\alpha}$  \\
\hline
12 & 1.5 & 100. & 5.0 \\
13 & 1.5 & 1000. & 5.1 \\
14 & 1.5 & 10000. & 4.6 \\
15 & 1.5 & 100000. & 3.8 \\
16 & 5. & 0. & 3.7-4.7 \\
17 & 10. & 0. & 2.9-4.4 \\
\hline
\end{tabular}
\end{center}
\end{table*}

\section{The \bp\ disk}
\label{bp disk}

\subsection{Orbiting-Evaporating-Bodies (OEB). Towards a solution?}

The observed distribution in the external part of the \bp\ disk ($\alpha \le
4.3$) is less steep than all calculated slopes in Tables~\ref{resultat_1}. It
has to be noted that a solution compatible with the observed slope could be
obtained with a gas density $\rho_0 \sim 10^{5}$~cm$^{-3}$ at 100~AU. However,
this is excluded by the upper limit deduced from spectroscopic or HI
observations (Vidal-Madjar et al. 1986, Freudling et al. 1995).

\begin{figure}[tbp]
\ifthisfig
\psfig{file=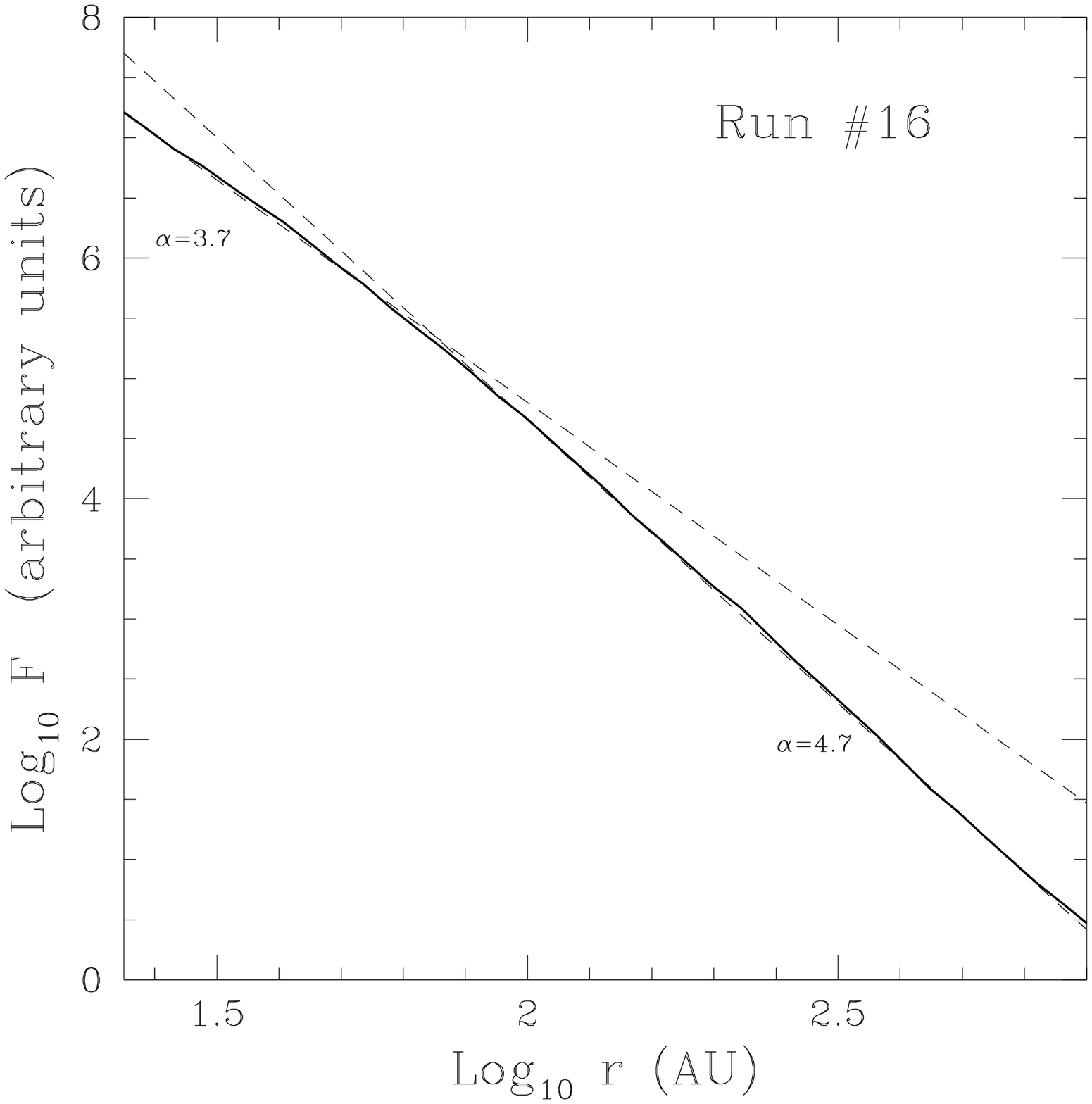,height=\columnwidth,rheight=8.5truecm}
\else\picplace{6truecm}\setfigdefault\fi
\caption[]{Plot of the surface brightness (arbitrary unit) of the equatorial
plane of an edge-on
disk seen from the Earth
calculated in run \# 16. The slope varies from -3.7 to -4.7.
For comparison, the dashed lines represent the power laws $r^{-3.7}$
and  $r^{-4.7}$.}
\label{16}
\end{figure}

Runs \#16 and \#17 are also compatible with the dust distribution in \bp\
(Fig.~\ref{16}). But they require dust size distributions ($K=5-10$) very
different from the expected ones if they are produced
through collisions between parent
bodies (Fujiwara 1979 and references therein). However they show us that we
need a relatively larger amount of smaller particles which have larger
apoastrons.

\begin{figure}[tbp]
\ifthisfig
\psfig{file=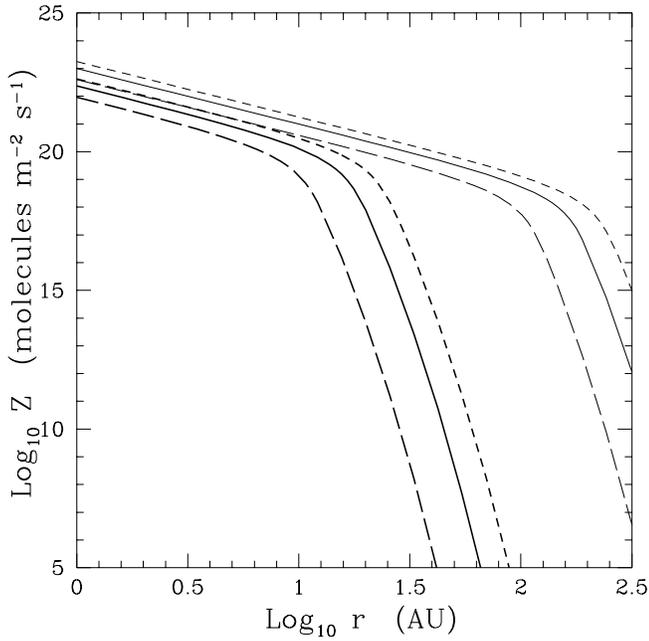,height=\columnwidth,rheight=8.5truecm}
\else\picplace{6truecm}\setfigdefault\fi
\caption[]{Plot of the evaporation rate $Z$ as a function of the distance
from \bp. The thick lines are for CO$_2$ and the thin ones for CO.
The evaporation
is calculated for a steady-state energy balance and depends on the albedo
$A_v$ of the parent body. The short-dashed lines are for $A_v$=0.1, the
long-dashed are for $A_v$=0.8 and the solid line are for $A_v$=0.5. In all the
simulations we take $A_v$=0.5.}
\label{Z}
\end{figure}

\begin{figure}[tbp]
\ifthisfig
\psfig{file=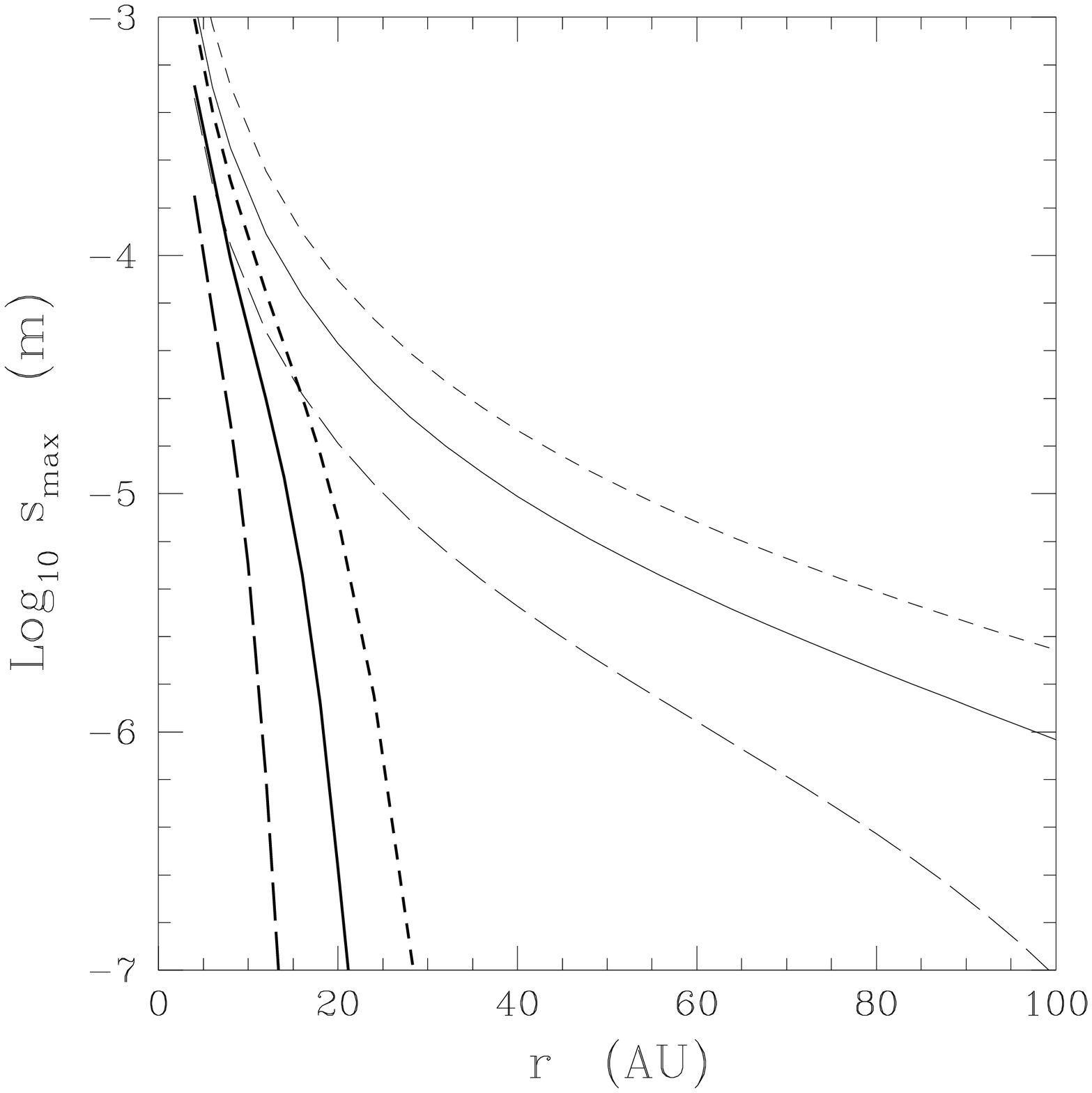,height=\columnwidth,rheight=8.5truecm}
\else\picplace{6truecm}\setfigdefault\fi
\caption[]{Plot of the radius of the largest particle which can be ejected
from the surface of a parent body with radius $R_{body}$=100~km and
density $\rho_{body}=1$
as a function of \bp\ distance. The thick lines are for CO$_2$ and the thin
ones
for CO. The different albedos $A_v$ are the same as in Fig.~3.}
\label{smax}
\end{figure}

In the evaporation process of large bodies (like Chiron), there is
an upper limit $s_{max}$ to the size of the ejected particles extracted by
gas (Luu \& Jewitt 1990). Following Cowan \& A'Hearn (1982) we have calculated
this upper limit and the evaporation rate $Z$ in molecules per
second per unit area for CO and CO$_2$ as a function
of the distance to \bp, assuming $L_{\beta Pic}=6$L$_{\odot}$
and a typical parent body
radius $R_{body}=100$~km (Figs.~\ref{Z} and \ref{smax}). This upper limit is
inversely proportional to the parent body size $R_{body}$.

We have introduced this effect in our model. The results are summarized in
Table~\ref{resultat_co2} for CO$_2$; they would be similar for CO if the parent
bodies were at larger distances. Here, we have taken the production rate
$p\propto Z\cdot R_{body}^2$
and a parent bodies size distribution
$dn=R_{body}^{-\gamma}dR_{body}$ where $\gamma=3.5$ (uncertainties
deduced from observations in the Solar System with $\gamma$
between 3.2 (Whipple 1975, Hughes \& Daniels 1982) and 3.8 (Fern\'andez 1982)
do not change the present results).

In runs \# 101 to \# 103, we take a wide belt of parent bodies from
15 to 30 AU, without correlation between their radius $R_{body}$ and their
distance from the central star. We
obtain $\alpha \sim 4.8-4.9$; the effect of cutoff in particle size
is not efficient
enough to explain the observed slope $\alpha \sim 3.6-4.3$.

However, if the evaporating bodies are coming from larger distances and their
orbital parameters diffuse from a Kuiper belt-like zone towards a
planetary-like
zone, the smaller parent bodies lost their volatile material very early at
larger
distances whereas the larger bodies can evaporate downward at
smaller distances.

This scenario can be modelized in the following way: a parent body
with a characteristic distance (which can be the periastron or the
semi-major axis) $r(t)=r_0-\dot{r}t$ is considered as evaporating at
$r=r(T)$ only if the total mass of the previously evaporated
gas is smaller
than the available mass of volatile material.
\begin{center}
\begin{equation}
\label{dead comet}
\int_{0}^{T} \mu Z(r(t))4\pi R_{body}^{2}\lambda dt < \frac{4\pi }{3}\rho \xi
R_{body}^3
\end{equation}
\end{center}

where $\lambda$ is the percentage of active surface of the body, $\xi$ the
relative mass of CO$_2$ in the parent body, $\rho$ its
density and $\mu$
the CO$_2$ molecular weight.
Eq.~\ref{dead comet} can be reduced to
$R_{body}\ge R_{ref}\int_{r}^{\infty}Z(l)dl/\int_{r_{ref}}^{\infty}Z(l)dl$
 where
$R_{ref}$ is the reference size of the smallest body which can evaporate at
$r=r_{ref}$. For CO$_{2}$ evaporation and $r_{ref}=20$~AU, one obtain
\begin{center}
\begin{equation}
\dot{r}=6\cdot 10^{-7} \frac{\lambda}{\xi \rho}
\left( \frac{10 {\rm km}} {R_{ref}} \right){\rm AU\ year}^{-1}
\end{equation}
\end{center}

\begin{figure}[tbp]
\ifthisfig
\psfig{file=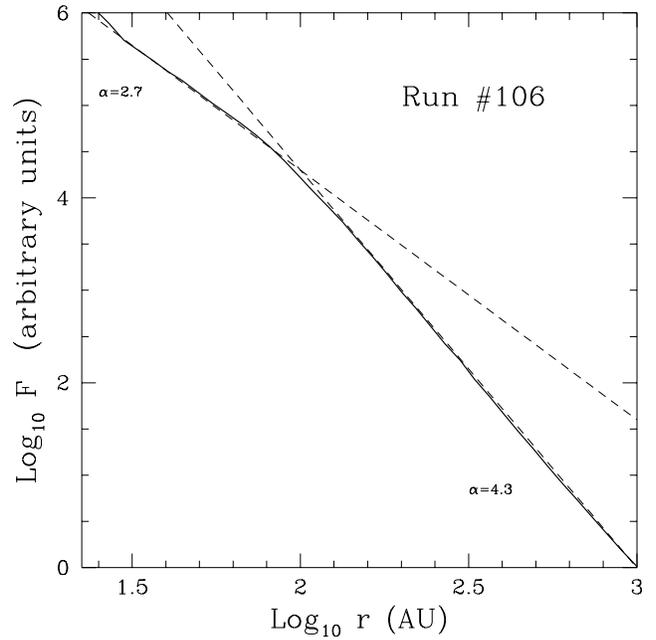,height=\columnwidth,rheight=8.5truecm}
\else\picplace{6truecm}\setfigdefault\fi
\caption[]{Plot of the surface brightness $F(r)$
calculated from run \# 106. There
is an abrupt change in the slope at $r=100$~AU. The slope is
$\alpha=-2.7$ inside and $\alpha=-4.3$ outside this limit.
For comparison, the dashed lines represent the power laws $r^{-2.7}$
and  $r^{-4.3}$.}
\label{106}
\end{figure}

In runs \# 104 to \# 115, only those with $R_{ref}\sim 30-40$~km give the
observed slope range. Moreover, due to the particle size cutoff,
$\beta \geq 0.4$ there is an abrupt break
in the slope at
$r_{break}\approx a_0/(1-2\beta_{min}) \approx 100\ {\rm AU}$
(see Fig.~\ref{106} for run \#106). For CO, one can obtain
$\dot{r}=6\cdot 10^{-6} \lambda /(\xi \rho) {\rm AU\ year}^{-1}$, and there
would be also an abrupt break in the slope since CO begins to evaporate
from about 100-150 AU.

\subsection{Discussion}

We saw in the previous section that the observed gradient in the \bp\ disk can
be
explained by the present model only if there is a larger amount of small
particles. Now, there is an upper limit in the size of particles ejected
by slowly evaporating bodies due to a balance between gas drag and
gravitation. In fact, we obtain dust disks
very similar to the \bp\ one by assuming evaporation
of kilometer size bodies which are diffusing from a Kuiper belt-like zone
inward an evaporation zone.

The time scale $\dot{r}\sim 10^{-7}$~AU~year$^{-1}$ in
the diffusion of comet orbital elements, which is necessarily to explain \bp's
characteristics, is
consistent with the few AU in 10 Myr deduced by Torbett \& Smoluchowski (1990)
and Levison's (1991) results on the Kuiper belt dynamical behavior between 30
and 40 AU around the Sun.
The analysis of the evolution of the orbital elements of the parent bodies
is outside the scope of this paper; moreover,
the motion of such Kuiper belt-like bodies under the influence of
a planetary system is probably chaotic (Scholl 1979, Torbett \& Smoluchowski
1990) and very dependent on this planetary system.
The present model merely shows us that
the dust distribution may be an important clue and may
be explained by the present model if there is
a long time scale perturbation of parent bodies by a planetary system which is
compatible with the one predicted for the Solar System.

Another evolution process for heating the bodies consists in the increase
of the star luminosity ($L$)
during its evolution along the main sequence. For a star of
1.6~M$_{\odot}$, we have $dL/Ldt=2\cdot 10^{-10}$~year$^{-1}$, if $\tilde{r}$
is the distance of equal luminosity, this
corresponds to $\dot{\tilde{r}}=\tilde{r}/2L\cdot dL/dt$.
This solution would have the advantage
of being "planetary system" independent. However, if the primary volatile is
CO,
$\dot{\tilde{r}}=10^{-8}$~AU~year$^{-1}$; even if the
volatile is CO$_2$, $\dot{\tilde{r}}=2.5\cdot 10^{-9}$~AU~year$^{-1}$.
The time scales do not seem to be adequate. This mechanism would be efficient
only if we could have bodies with very large quantities of volatiles and
very small fractions of active surface.

The assumption that CO or CO$_2$ can be primary volatiles is not a problem
since around an AV star they can evaporate at very large
distances where the
presence of young objects without a lag deposit of nonvolatile forming a crust
is likely. On the contrary in the Solar System, we observe bright
comets only when H$_2$O can evaporate, simply because
CO or CO$_2$ evaporate only inside
planetary regions where, for dynamical reasons, the presence of young objects
is unlikely: Chiron seems to be an exception but provides a useful analogy
with what may happen around \bp.

As we shall see below, this model is also able to give answers
to other important issues defined in Section~\ref{intro}.

\begin{table*}
\caption{Same as the previous tables, for particles ejected by evaporating
gas. Here we assume that the gas is CO$_2$ and $p\propto Z\cdot R_{body}^2$.
$R_{ref}$ is the radius of the largest dead
comets at $r_{ref}$=20~AU. For runs \# 105 to \# 115
there is a slope break at $r=r_{break}$, and both inward and outward slopes
$\alpha _{in}$ and
$\alpha _{out}$ are given. For runs
\# 113 to \# 115 the two given slopes $\alpha _{in}$ are for each side
of the disk. We always took $t_{max}$=10000 years.}
\label{resultat_co2}
\begin{tabular}{|c|ccc|c||ccc|c|}
\hline
Run && Parent Bodies Parameters  &&
$R_{ref}$& &Result: & $F(r)\propto r^{-\alpha}$  & Brightness\\
&&&&  & $r_{break}$ & $\alpha _{in}$ &
$\alpha_{out}$ & ratio\\
$\#$ &  Periastron (AU) & Eccentricity & $\omega$ & (km) &(AU)&&&\\
\hline
101 & 20-30 & 0.       & -         &   -  & -   & - & 4.8 & - \\
102 & 20-30 & 0.-0.5   & 0-360$^o$ &   -  & -   & - & 4.9 & - \\
103 & 20-30 & 0.-0.5   & 155$^o$    &   -  & -   & - & 4.9 & - \\
104 & 15-30 & 0.       & -         & 10.  & -  & -  & 4.6 & - \\
105 & 15-30 & 0.       & -         & 20. & 50  & 2.9 & 4.5 & - \\
106 & 15-30 & 0.       & -         & 30.  & 100 & 2.7 & 4.3 & - \\
107 & 15-30 & 0.       & -         & 40. & 180 & 2.5 & 3.9 & - \\
108 & 15-30 & 0.-0.5   & 0-360$^o$ & 10.  & -  & -  & 4.6 & - \\
109 & 15-30 & 0.-0.5   & 0-360$^o$ & 20. & 50  & 2.8 & 4.5 & - \\
110 & 15-30 & 0.-0.5   & 0-360$^o$ & 30.  &  130 & 2.8 & 4.3 & -  \\
111 & 15-30 & 0.-0.5   & 0-360$^o$ & 40.  &  180 & 2.6 & 3.7 & - \\
112 & 15-30 & 0.-0.5   & 155$^o$    & 10.  &  -  & - & 4.6 & 2.9 \\
113 & 15-30 & 0.-0.5   & 155$^o$    & 20. &  50  & 2.2-4.0 & 4.5 & 2.8 \\
114 & 15-30 & 0.-0.5   & 155$^o$    & 30.  &  110 & 2.7-3.7 & 4.3 & 3.5 \\
115 & 15-30 & 0.-0.5   & 155$^o$    & 40.  &  180 & 2.7-3.4 & 3.7 & 4.0 \\
\hline
\end{tabular}
\end{table*}

\subsection{Asymmetry}

The distribution observed at large distances in \bp\ disk
is obviously not axisymmetric
(Smith \& Terrile 1987, Kallas \& Jewitt 1995).
However,
relative keplerian motions should remove such asymmetries: for two particles
on circular orbits with semi-major axis $a_1$ and $a_2$, the time needed
to put them in opposite side is $t\sim 0.4(a_1^{-3/2}-a_2^{-3/2})^{-1}$~years
if $a_1$ and $a_2$ are in AU. With $a_1=100$~AU and $a_2\ge400$~AU we have
$t\le 450 $~years, an extremely short time compared to the age
of the system. Thus,
any asymmetry should quickly disappear.

\begin{figure}[tbp]
\ifthisfig
\psfig{file=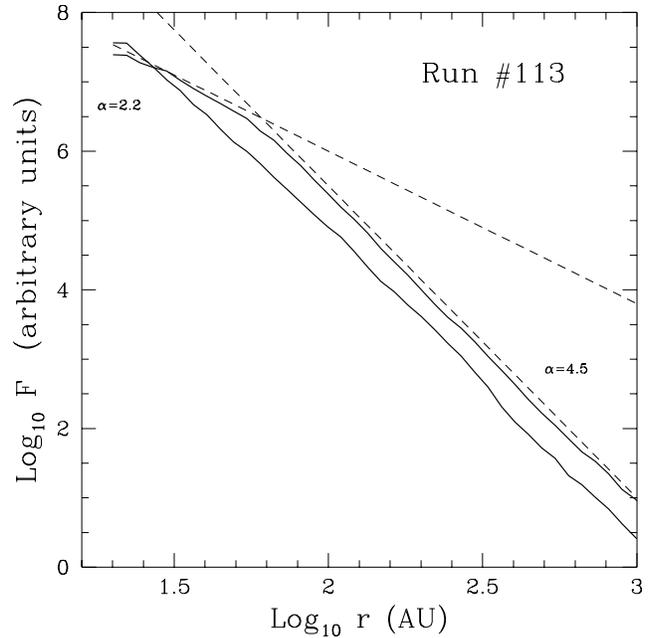,height=\columnwidth,rheight=8.5truecm}
\else\picplace{6truecm}\setfigdefault\fi
\caption[]{Plot of the surface brightness of the two extension
calculated from run \# 113.
In this run all the parent bodies have a longitude of periastron
of $155^{o}$ relative to the line of sight. The brightness ratio between
the two extensions is equal to 2.8.
For comparison, the dashed lines represent the power laws $r^{-2.2}$
and  $r^{-4.5}$.}
\label{113}
\end{figure}

As noted above, it is however
possible to find the asymmetry observed between the two
extensions of the
\bp\ disk by assuming a peculiar parent body distribution (see
Fig.~\ref{113}).
Runs
\# 112 to \# 115
have been obtained with a peculiar longitude of periastron of
the parent bodies ($\omega=155^o$).
We found a brightness
ratio between the two extensions of the disk ranging from 2.8 to 4.0.
This result shows that
the observed ratio of SW to NE extension brightness from 1.1
(Lecavelier des Etangs et al. 1993) to 1.5 (Kallas \& Jewitt 1995) can be
reproduced if there is a small additional proportion of bodies
with peculiar longitudes of periastron like in the FEB scenario.

\begin{figure}[tbp]
\ifthisfig
\psfig{file=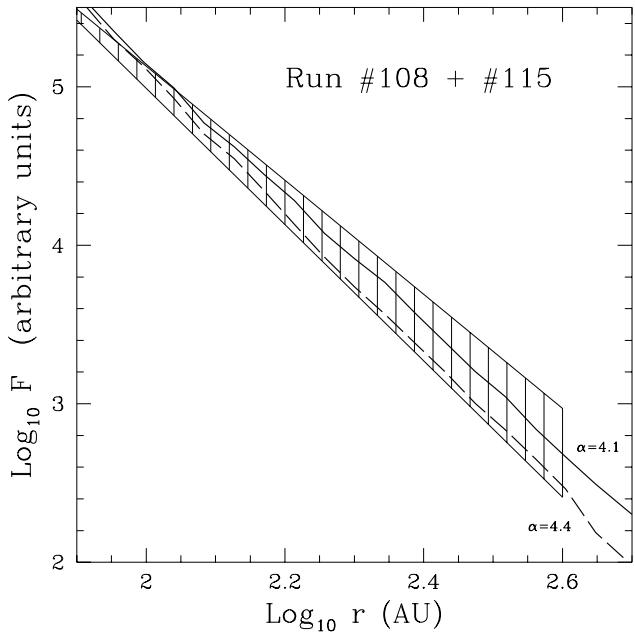,height=\columnwidth,rheight=8.5truecm}
\else\picplace{6truecm}\setfigdefault\fi
\caption[]{Plot of the surface brightness calculated from the addition
of runs \# 108 and \#115.
In this combined run the parent bodies with a longitude of periastron
of $155^{o}$ go more slowly inwards. The brightness ratio between
the two extensions is equal to 1.3. The fainter extension has a steeper slope
and is represented by a long-dashed line.\\
As in Fig.~1., the hatched zone represents the distribution range of slopes
deduced by different observers around \bp.}
\label{108+115}
\end{figure}

However, there is another way to obtain the observed asymmetry, namely
to assume that bodies
with different longitudes of periastron move inwards with different velocities.
For example, if bodies with a given longitude of periastron move more slowly
inward, it produces a
fainter disk extension with a steeper brightness gradient in one particular
direction (Fig.~\ref{108+115}).

\subsection{Mass of the disk}
\label{mass}

The total mass of the \bp\ disk calculated from optical observations is
\begin{center}
\begin{equation}
M=\int\int n_s(r,s) \frac{4\pi }{3}\rho s^3 2\pi r dr dn_s
\end{equation}
\end{center}

where $s$ is the particle size and $dn_s$ the size distribution. As discussed
above we can take $dn_s=s^{-3.5}ds$. Previously, authors assumed that
$s$ is independent of $r$ the distance from the star. This assumption
is the simpler one and others must be justified.

This means however that:
\begin{center}
\begin{equation}
M\propto
\frac{8\pi ^{2}\rho }{3} \int_{r_{in}}^{r_{out}}n_s(r)rdr \int s^{-0.5}ds
\end{equation}
\end{center}

The second integral diverges towards the largest bodies; the first
one also diverges for $r_{out} \rightarrow  \infty$ if the slope of the
scattering light is $\ge-4$ (that is the case according to Artymowicz et al.
(1989), Lecavelier des Etangs et al. (1993) and for
Golimowski et al.'s (1993) NE
extension). Thus to evaluate the mass, an arbitrary limit for the disk
extension and particle size must be fixed.

In the framework
of our model both problems are directly solved. At a given distance
$r$, we have
$
2\epsilon \le s \le \frac{2\epsilon r}{r-a_0}
$
where $a_0$ is the distance of the parent body to the star and $\epsilon$
is defined by $\epsilon=s\beta $. We saw in Section~\ref{size} that
$\epsilon$ can be assumed to be constant. We obtain
\begin{center}
\begin{equation}
M \propto \frac{4\pi \rho}{3}\int \left( n_s(r)2\pi r dr
\int_{2\epsilon}^{(2\epsilon r)/(r-a_0)}s^{-0.5}ds\right)
\end{equation}
\end{center}

\begin{center}
\begin{equation}
M \propto \frac{16\pi ^2}{3}\rho \sqrt{2\epsilon}
\int_{r_{in}}^{\infty} n_s(r)\frac{a_0}{2}dr
\end{equation}
\end{center}

This integral now converges and the computed mass
is of the order of few lunar masses, i.e.
of the same
order as the lower limit evaluated by Artymowicz et al. (1989) who assumed
an arbitrary outer limit of 500~AU and single-size particles
of $1\mu$. This mass is consistent with the total
mass deduced from submillimeter observations (Zuckerman \& Becklin 1993).

\subsection{Gas-dust ratio and CO detection.}
\label{CO}

One can investigate the connection between the dust observed
around \bp\ since 1984 and the gas disk with its stable component thanks to
the recent detection of CO in UV lines with the HST
($N_{co}\approx 10^{15}$~cm$^{-2}$, Vidal-Madjar et al. 1994).
The very presence of CO may also need a permanent replenishment naturally
provided by evaporation of comet-like bodies.

We can evaluate the total mass of dust $M_d$ which is associated with the
evaporation of CO:
\begin{center}
\begin{equation}
M_d=M_{co} \varphi \psi _{dust/gas} \psi _{gas/CO}  \frac{t_d}{t_{CO}}
\end{equation}
\end{center}
where $M_{CO}$ is the total mass of CO in the disk, $\psi_{dust/gas}$ and
$ \psi_{gas/CO}$ are the mass ratios of dust to gas and gas to CO, $\varphi$ is
the mass ratio of the dust produced effectively kept in the disk and
$t_d$ and $t_{CO}$ are the life time of dust and CO: due to photodissociation
by UV interstellar radiation $t_{CO}\sim 300$~years (Vidal-Madjar et al. 1994),
and
at a distance of 100 AU where CO begins to be vaporized, $t_d \sim 10^6$~years
(Backman \& Paresce 1993).
{}From the upper limit to particle size ($\beta \ge 0.4$), one can evaluate
$\varphi\sim 0.1$.
With a simple geometry of a disk with an opening angle of 10$^o$,
we can deduce from the observed CO column density
a total CO mass of about
$M_{co} \approx 2\cdot 10^{20}$~kg.
{}From observations in the Solar System, we can take $\psi_{dust/gas} \ge 0.1$
(Newburn \& Spinrad 1989). Finally, $\psi_{gas/CO}$ depends on the composition
of the
evaporated gas; it is between 1. if only CO is present
and $\sim 10.$ if all volatiles evaporate (Mumma et al. 1993).
Thus, with $\varphi \psi _{dust/gas} \psi _{gas/CO}  \approx 0.1$,
$M_d$ is about one lunar mass, that is of the same
order as the total mass of the dust disk independently evaluated!

This rough calculation shows that CO and dust are compatible with a common
origin
and produced by the same process: evaporation of comet-like bodies at large
distances from the star.

\section{Conclusion}
\label{conclusion}

The model we have proposed is able to account for the main characteristics of
the \bp\ disk as natural consequences of the production of secondary origin
particles by small bodies. It is in fact an answer to the question 3 of
Zuckerman \& Becklin (1993): are
the dust grains primordial or continually replenished?

Indeed, we have shown that if bodies produce dust particles
with a given size distribution, these particles follow very eccentric orbits.
Thus, a disk can be seen
at large distances from the star, and the spatial distribution of dust
is very close to a power-law.
Moreover, the particles have the same inclinations as
the parent bodies: the thickness of the disk thus
increases with radius, and the
disk is a "wedge" disk: (issue 4).
The asymmetric spatial distribution can be explained if we assume that
the parent bodies distribution is not axisymmetric (issue 2).
This assumption is not surprising since asymmetry is observed for the FEBs
and is probably the case in the Solar System for Kuiper belt objects trapped
in planetary resonances (Jewitt \& Luu 1995).
Since particles are produced with periastrons greater than or equal to those of
the parent
bodies, the dust present close to the star is either produced there or
brought in by Poynting-Robertson drag: the central part must be relatively
clear (issue 3).
Finally, this model gives connection between the particle size and their
distances from the central star, and enable us to solve the issue
of the disk mass (issue a), Section~\ref{mass}).
Thus, the issues 2), 3), 4) and a) listed in Section~\ref{intro} are
simultaneously solved with the simplest form of the model presented here.

However, in order to explain the flatter gradient observed in the \bp\ disk
than the one obtained without additional hypothesis, we propose to add two
other assumptions:\\
1) The dust is produced by evaporation of bodies of few kilometers in radius,
so that there is a balance between evaporating
gas drag and gravitation of these parent bodies,
and only the smallest particles are extracted.\\
2) These bodies are moving slowly towards \bp, so that they become extinct
before arriving close to the star where the evaporating rate is large enough
to produce the largest particles. \\
This OEB (Orbiting-Evaporating-Bodies)
scenario is proposed to solve issue 1), but surprisingly
it is also able to explain the abrupt change in the slope by the
cutoff in distribution of particle size (issue 5)). It has to be noted that
the cutoff must not necessary be very sharp and in fact is rather smooth
in the model since we assumed a parent body size distribution.
The model gives also a natural
connection between the dust and gas disks (issue b): they may well be produced
by the same bodies in the same process.
The connection between the FEB and asymmetry in the disk seems to be possible
(issue c)), however this point needs further analysis.

This new model is able for the first time to explain simultaneously
these issues, but we must now ask if other models or hypothesis can also
be made with the same consequences. If correct, this OEB scenario allows
to see the \bp\ disk as a gigantic multi-cometary tail with all its
components: gas and dust.

\begin{acknowledgements}
We would like to express our gratitude to
Dr. F.Roques who first recall us that ejected particles follow
peculiar eccentric orbit.
We are particularly indebted to the anonymous referee for his very
useful comments
which substantially improved the paper.
Our thanks go also to Dr. M. Friedjung for improving the
manuscript.
\end{acknowledgements}

\end{document}